\documentclass[aps,prl,twocolumn,superscriptaddress, showpacs]{revtex4}
\usepackage{graphicx}

\begin{document}

\preprint{submitted to PRL}


\title{Magnetic Reconnection by a Self-Retreating X-Line}


\author{M. Oka}
\email[]{mitsuo.oka@uah.edu}
\affiliation{Center for Space Plasma and Aeronomic Research, University of Alabama in Huntsville, Alabama, 35805, USA}

\author{M. Fujimoto}
\affiliation{Institute of Space and Astronautical Science, Japan Aerospace Exploration Agency, Sagamihara, Japan}

\author{T. K. M. Nakamura}
\affiliation{Institute of Space and Astronautical Science, Japan Aerospace Exploration Agency, Sagamihara, Japan}

\author{I. Shinohara}
\affiliation{Institute of Space and Astronautical Science, Japan Aerospace Exploration Agency, Sagamihara, Japan}

\author{K.-I. Nishikawa\altaffiliation{center}}
\affiliation{Center for Space Plasma and Aeronomic Research, University of Alabama in Huntsville, Alabama, 35805, USA}


\date{\today}

\begin{abstract}
Particle-in-cell (PIC) simulations of collisionless magnetic reconnection are performed to study asymmetric reconnection in which an outflow is blocked by a hard wall while leaving sufficiently large room for the outflow of the opposite direction. This condition leads to a slow, roughly constant motion of the diffusion region away from the wall, the so-called `X-line retreat'. The typical retreat speed is $\sim$0.1 times the Alfv\'{e}n speed. At the diffusion region, ion flow pattern shows strong asymmetry and the ion stagnation point and the X-line are not collocated. A surprise, however, is that the reconnection rate remains the same unaffected by the retreat motion.

\end{abstract}


\maketitle

Magnetic reconnection triggers many explosive phenomena in laboratory and astrophysical plasma. Importance lies in the diffusion region where the MHD breaks down and the kinetic scale comes in. It is a scientific challenge to understand the structure of the diffusion region and its role in controlling the reconnection rate. Owing to the growing power of computer capabilities, full particle simulations has become popular these days. It is now clear that the out-of-plane current remains localized and the reconnection rate remains fast while electrons form a high-velocity jet that extends large distances downstream from the X-line \citep{daughton06, fujimoto06, shay07, karimabadi07}. The region with the localized current is termed `inner diffusion region' whereas the rest of the diffusion region including the elongated electron current layer is termed `outer diffusion region'. In general, simulations generate magnetic reconnection that is symmetric in both inflow and outflow directions.

Actual magnetic reconnection, however, occurs in a more complicated situation. During a magnetic reconnection in the Earth's magneto-tail, an earthward outflow soon collides with the dipole field while the tailward outflow is directed to the interplanetary space. As for the solar flares that take place above the area of emerging fluxrope, the upward outflow eventually merge into the solar wind while the downward outflow collides with the magnetic obstacle in the lower part of the corona. In these cases, magnetic reconnection may not be symmetric about the X-line.

In this letter, we investigate the consequences of blocking one side of the outflow. A hard wall is set up just ahead of an outflow while leaving a sufficiently large room for the outflow of the opposite direction. As a result, reconnected magnetic fields are piled-up against the wall. By the time the pile-up region reaches the diffusion region, the X-line starts to move away from the wall (`X-line retreat'). Under such a strong influence of the boundary condition, the structure of the diffusion region is modulated. In the past, the X-line motion was considered by an analytical treatment \citep{owen87}, MHD simulations \citep{priest02}, and a full particle simulation \citep{hesse01}, although these work did not shed light on the modulated structure of the diffusion region. This paper will be the first to report the structure of the diffusion region in a self-consistent simulation of asymmetric outflow reconnection.

We utilized the two dimensional, particle-in-cell code \citep{hoshino87, shinohara01}. The initial condition consists of a Harris current sheet. The anti-parallel magnetic field and the density are given by $B_y=B_0\tanh((x-L_x/2)/D)$ and $N_{cs}=N_0/\cosh^2((x-L_x/2)/D)$, respectively, where $B_0$ is the magnetic field at the inflow boundary, $D$ is the half-thickness of the current sheet, $N_0$ is the density at the current sheet center and $L_x$ and $L_y$ are the domain size in $\mathbf{\hat{x}}$ and $\mathbf{\hat{y}}$ direction, respectively. $D$ is chosen to be 1.5$\lambda_i$ where $\lambda_i$ is the ion inertial length. The inflow, background plasma is represented by $N_{B}=N_{B0}(1-1/\cosh^2((x-L_x/2)/D))$ where $N_{B0}$=0.2$N_0$. The pressure imbalance by the non-uniform density is not important to the results presented below. The electron to ion temperature ratio is set to be $T_e/T_i$=1/5 for both the current sheet and the background. The frequency ratio $\omega_{pe}/\Omega_{ce}$=1.5 where $\omega_{pe}$ and $\Omega_{ce}$ are the electron plasma frequency and the electron cyclotron frequency, respectively. The conducting walls are used at $x$=0 and $x$=$L_x$ and symmetric boundary conditions were used at $y$=0 and $y$=$L_y$. At the symmetric boundary, particles are specularly reflected whereas fields are given by $\partial B_n/\partial n$ = 0, $\mathbf{B_t}=0$, $\partial \mathbf{E_t}/\partial n$ = 0, and ${E_n}=0$, where $\mathbf{E}$ and $\mathbf{B}$ are the electric field and the magnetic field vectors and the subscripts n and t denote the normal and tangential components, respectively. Reconnection is initiated at the distance $H$ from the bottom ($y$=0) wall with a small magnetic island given by the vector potential $A_z=-A_0\exp[-\{(x-L_x/2)^2+(y-H)^2\}/(2\lambda_i^2)]$ where $A_0=0.1B_0\lambda_i$. $L_y$ is chosen to be large enough to simulate effectively the free boundary condition of the reality.

Five simulation runs are performed to study the dependence on the initial distance $H$. The values of $H$ along with the other initial values are compiled in Table \ref{tab:table}. The initial X-line is at the center of the domain for Run 4, and thus, it is a non-retreating case. Run 5 is an additional case intended to study the dependence on the ion to electron mass ratio $\mu\equiv m_i/m_e$. In all runs, we used average of 64 particles in each grid cell. 276 particles per cell represents the unit density.

\begin{table}
\caption{Initial parameters and results of each run. Lengths are normalized by the ion inertial length $\lambda_i$, times are by the inverse ion cyclotron frequency $\Omega_{ci}^{-1}$ and speeds are by the ion Alfv\'{e}n speed $v_A$.  $\Delta$ is the grid size. See text for details.\label{tab:table}}
\begin{ruledtabular}
\begin{tabular}{ccccccccc}
& $L_x$ & $L_y$     & $H$ & $\mu$ & 
$\lambda_i/\Delta$ & $t_{col}$ & $t_{ret}$ & $v_{p}$\\ \hline
Run 1&  76.8 & 204.8 &  12.8 & 25 & 20 &  80 &  83 & 0.3 \\
Run 2&  76.8 & 204.8 &  25.6 & 25 & 20 &  90 &  98 & 0.5 \\
Run 3& 102.4 & 204.8 &  51.2 & 25 & 20 & 110 & 140 & 1.0 \\
Run 4& 102.4 & 204.8 & 102.4 & 25 & 20 & 120 &  -  &  -  \\
Run 5& 170.7 & 341.3 &  25.6 &  9 & 12 &  90 &  98 & 0.5 \\
\end{tabular}
\end{ruledtabular}
\end{table}


\begin{figure}[b]
\includegraphics[width=80mm]{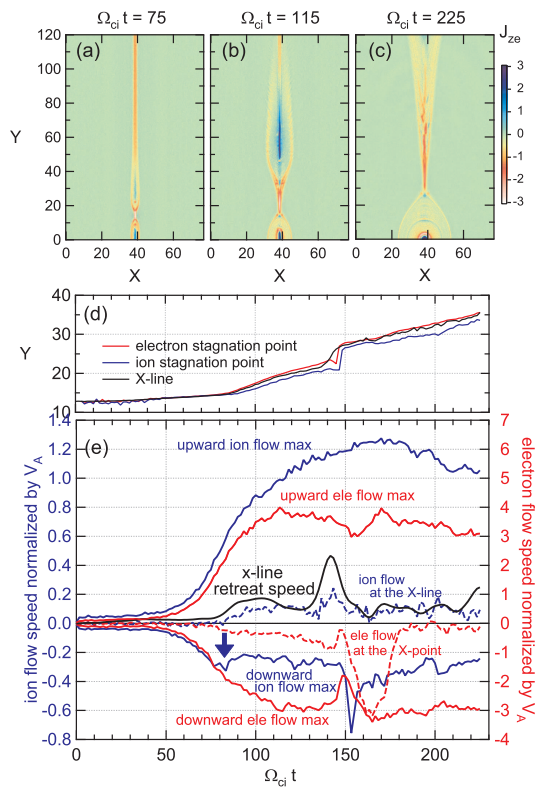}
\caption{Overview of the retreating X-line for Run 1: (a-c) normal electron current $J_{ez}$. (d) positions of the X-line and the stagnation points. (e) the X-line retreat speed (black curve), upward and downward maximum flow speeds for ions (blue solid curves) and electrons (red solid curves), and the flow speeds at the X-line for ions (blue dashed curve) and electrons (red dashed curve). A stagnation point is defined by the flow reversal point. \label{fig:retreat}}
\end{figure}


Upper panels of Figure \ref{fig:retreat} show snapshots of electron out-of-plane current for Run 1. It is evident that as magnetic reconnection proceeds, a downward flow is blocked by the bottom wall and creates the pile-up region which expands in time and eventually pushes the diffusion region upward. The upward flow, on the other hand, continues to grow in length until its leading edge reaches the upper wall. The choice of $L_y$ is not sensitive to the results presented in this paper and thus any effect from the upper wall is not important. The diffusion region is also elongated and the final length of the electron current sheet becomes as long as $\sim$50$\lambda_i$.  

The black curve in Figure \ref{fig:retreat}d shows the X-line position defined by the $B_x$ reversal point, $B_x$=0. A distinct retreat process starts at $\Omega_{ci}t\sim$85. The X-line moves away from the wall monotonically. Although a jump appears at $\Omega_{ci}t\sim$145 due to a magnetic island generated by a secondary tearing instability, the X-line motion continues even after the upward outflow reaches the upper wall. The approximate time of the outflow collision with the wall is $\Omega_{ci}t\sim$180.

Figure \ref{fig:retreat}e shows various flow speeds. While the upward ion flow becomes super-Alfv\'{e}nic and continues to increase until the time of collision with the wall, the downward ion flow reaches $\sim$0.3$v_A$ at $\Omega_{ci}t\sim$80 when it is blocked by the bottom wall (marked by the blue arrow). The downward flow speed soon decreases and is followed by the upward motion of the X-line. In contrast, the retreat speed and the ion flow speed at the X-line are roughly constant at the same value $\sim$0.1$v_A$, although disturbances appear at $\Omega_{ci}t\sim$145 due to a secondary tearing. 

\begin{figure*}
\includegraphics[width=160mm]{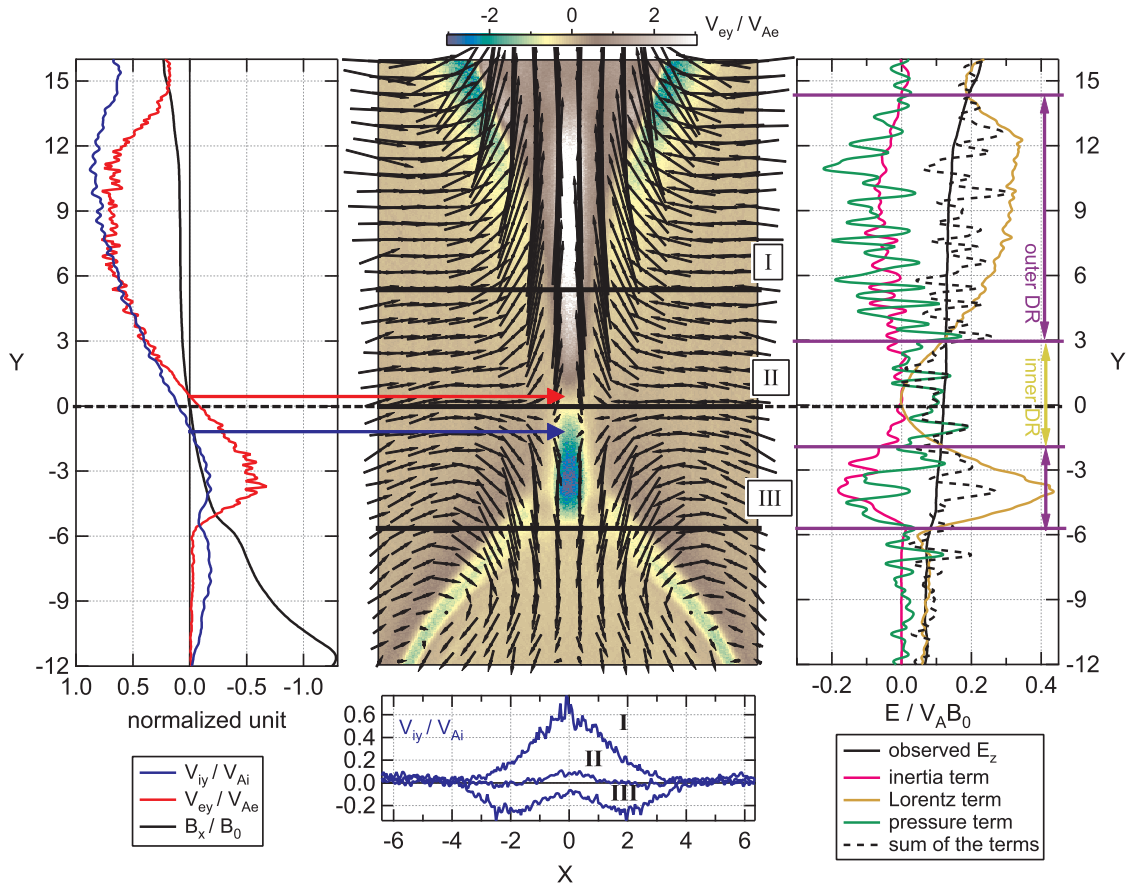}
\caption{Enlarged view of the retreating X-line for Run 1 obtained at $\Omega_{ci}$t=115. The center panel shows the electron flow in $\mathbf{\hat{y}}$ direction $V_{ey}$ (color code) as well as ion flow directions in the $x-y$ plane (vectors). The left panel shows quantities $B_x$, $V_{ey}/\sqrt{\mu}$, and $V_{iy}$ along $x$=0 whereas the right panel shows the electric field terms of the generalized Ohm's law along $x$=0. The bottom panel shows $V_{iy}$ cut along the horizontal lines in the color code. The purple and yellow arrows in the right panel show the outer and inner diffusion regions (DR), respectively.\label{fig:dif}}
\end{figure*}

A striking feature can be found around the diffusion region. Figure \ref{fig:dif} shows an enlarged view of the diffusion region for Run 1 at $\Omega_{ci}t\sim$115 when the upward electron speed is saturated. In this figure, data are accumulated over one gyro period. Coordinates are shifted so that the X-line comes to the origin. This simulation, along with others with different $H$, reveals that the ion and electron stagnation points are not collocated  with the X-line. As the pile-up region expands, the lower half of the outer diffusion region is modulated and ion flows are deflected. As a result, the ion stagnation point is shifted downward about the distance of $\lambda_i$ from the X-line (marked by the blue horizontal arrow). This separation between the stagnation point and the X-line is almost constant in time as shown in Figure \ref{fig:retreat}d. The shift of the ion stagnation point leads to an upward ion flow at the X-line, resulting in a slow, rising motion of the X-line. While there is a high speed ion flow in the upper half of the diffusion region (the curve labeled `I' in the bottom panel), the typical value of the upward ion flow at the X-line was $\sim$0.1$v_A$ (`II'). The deflection can be recognized as the double peak feature in the cut of the downward outflow (`III'). 

In contrast to ions, electrons do not show strong deflection and exhibit well developed outflow jets both upward and downward, although the downward jet is limited in length. The typical length of the downward electron jet is $\sim$5$\lambda_i$ and remains constant throughout the run. The maximum value of the electron flow speed, however, is quite fast and reaches 0.65$v_{Ae}$, as can be seen in the left panel. This is almost the same as the maximum speed of the upward electron jet. We have found that all retreating cases show nearly the same maximum speed of the downward electron jet, $i.e.$ 0.7$v_{Ae}$. The length of the jet, on the other hand, becomes slightly larger for larger $H$. For Run 3, the length reached $\sim$10$\lambda_i$.

We further examined the out-of-plane component of the generalized Ohm's law along $x$=0 as shown in the right panel of Figure \ref{fig:dif} \citep{shay07}. It is expressed as
\begin{equation}
E_z = -\frac{m_e v_{ey}}{e}\frac{\partial v_{ez}}{\partial y}+\frac{1}{c}v_{ey}B_x - \frac{1}{ne}\nabla\cdot\mathbf{\Gamma}
\end{equation}
where $\mathbf{v_e}$ is the electron bulk velocity, $\mathbf{\Gamma}=p_{exz}\mathbf{\hat{x}}+p_{eyz}\mathbf{\hat{y}}$ is the flux of $z$-directed electron momentum in the reconnection plane (not including convection of momentum) with $\mathbf{p_e}$ the electron pressure tensor. The observed electric field (solid black curve) is balanced by the sum (dashed black curve) of the electron inertia (red), the Lorentz force (brown), and the divergence of the momentum flux (green). Note that the time variation term is negligible and is not considered here. As is the case for non-moving, symmetric reconnection reported previously \citep{shay07}, the major contribution to the inner diffusion comes from the pressure term. The upward outer diffusion region is also similar to the symmetric case. On the other hand, the downward outer diffusion region is highly modified. Steeper acceleration of $v_{ey}$ and piled-up $B_x$ make the Lorentz term profile to be sharper. The inertia term with the opposite sign is also enhanced because of steeper gradient of $v_{ez}$ in the short jet. The striking feature is that both are enhanced in a balanced way such that $E_z$ as a whole shows the almost flat profile along the $x$-axis.

\begin{figure}[b]
\includegraphics[width=90mm]{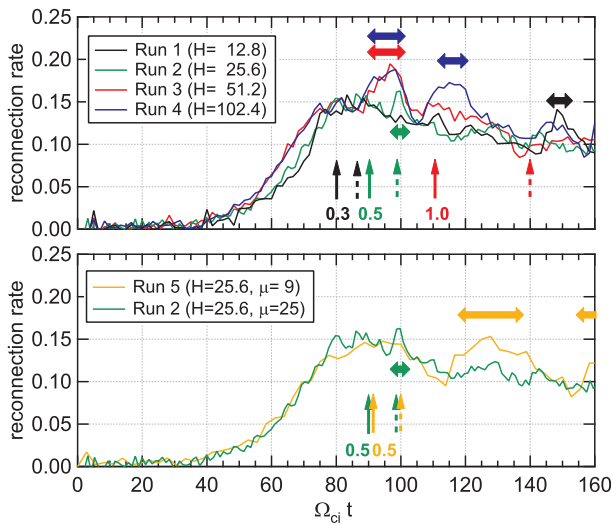}
\caption{Dependencies of reconnection rate on the initial distance from the wall (upper panel) and the mass ratio $\mu$ (lower panel).  The horizontal arrows show periods of secondary island appearances. The straight vertical arrows show the approximate times of the collision of the downward jet $t_{col}$ which is defined by the peak in the profiles of the downward ion flow speed (see, for example, the blue arrow in Figure \ref{fig:retreat}). The dashed vertical arrows show the start times of the X-line retreat, $t_{ret}$. The numbers annotated below each dashed arrow are the downward ion flow maximum $v_p/v_A$. There is no vertical arrow for Run 4 because it is the non-retreating case.
\label{fig:rxr}}
\end{figure}

The above finding leads to the idea that the downward outer diffusion region adjusts itself to buffer the effects of the wall such that the reconnection electric field is determined irrespective of the asymmetric boundary condition. This idea is inspected in Figure \ref{fig:rxr} which shows the reconnection rate, $E_{\rm R}$, of all runs. $E_{\rm R}$ is determined by averaging the electric field over a small square region centered at the X-line. We have verified that taking the time derivative of the total magnetic flux between the X-line and the bottom center of the pile-up region yields identical profile as $E_{\rm R}$. For Runs 1-3 and 5, the upward outflow jet reaches the upper wall well after $\Omega_{ci}t\sim$160. For Run 4, both upward and downward jet collide at around $\Omega_{ci}t\sim$120. It is evident that retreating cases (Runs 1-3) shows almost the same profile as the non-retreating case (Run 4). Moreover, the profile for Run 5 confirms the earlier result that $E_{\rm R}$ does not depend on $\mu$ \citep{hesse99}. Relatively large fluctuations are due to secondary islands whose appearance times are indicated by the horizontal arrows. 
Also illustrated in Figure \ref{fig:rxr} are the timings of the jet collision with the wall as well as the X-line retreat.  It clearly shows that the rising motion of the diffusion region is associated with the expansion of the pile-up region rather than the peak $E_{\rm R}$.

To summarize, the retreat motion of the X-line is constant ($\sim$0.1$v_A$). This speed does not depend on the initial distance of the X-line from the wall, $H$, except of course the non-retreating case. The downward outflow region is largely modulated by the asymmetry. The ion deflection pattern depends on $H$ and so is the length of the electron downward jet. The outer diffusion region is not elongated as in the case of symmetric reconnection. In terms of the reconnection rate, however, it stays the same as the symmetric case thanks to the internal balance within the downward outer diffusion region. 

In addition to the initial distance from the wall, a dependence on the mass ratio $\mu$ should also be addressed. What we have found here is the tendency for more magnetic islands to emerge with smaller $\mu$. These islands may help maintain fast reconnection \citep{daughton06}. In fact, as shown in Figure \ref{fig:rxr}, the $\mu$=9 case (Run 5) generated two large islands, yet keeping the same reconnection rate as the other runs. Another important aspect of the $\mu$ dependence is the separation between the ion stagnation point and the X-line, but we could not measure $\delta_{\rm xs}$ for the $\mu$=9 case because of the successive generation of the secondary islands. The mass dependence of the size of the separation will be studied in our future paper. Note the fact, however, that the retreat speed was roughly equal to the ion flow speed so that the X-line and the stagnation will be collocated in the X-line rest-frame of reference.

Finally, it should be mentioned that the X-line speed can be faster than the self-retreating speed of 0.1$v_A$ if, say, it is blown by another dominant X-line. Recent two-fluid simulations show that its reconnection rate is reduced and reconnection at the moving X-line is eventually terminated. Particle simulations for this kind of situations are also needed.

\begin{acknowledgments}
This work was partially supported by the Grant-in-Aid for Creative Scientific Research (17GS0208) from the MEXT, Japan. MO was supported by the Grant-in-Aid for JSPS Postdoctoral Fellows for Research Abroad.
\end{acknowledgments}


\end{document}